\documentstyle[twoside,fleqn,npb,epsfig]{article}
\newcommand{\AmS}{{\protect\the\textfont2
  A\kern-.1667em\lower.5ex\hbox{M}\kern-.125emS}}

\title{
Adler function from $R^{e^+e^-}(s)$ measurements:
experiments vs QCD theory.}

\author{A.L. Kataev\address{Institute for Nuclear Research of the
        Academy of Sciences of Russia,
        117312 Moscow, Russia}
        \thanks{
Supported
in part by
the Russian Foundation
        of Basic Research, Grant N 99-01-00091}
\thanks{Contributed
to the Proceedings of the International Workshop ``$e^+e^-$ collisions
from $\Phi$ to $J$/$\Psi$'', Novosibirsk, March 1-5 1999.}}
\begin{document}
\begin{abstract}
An experimentally motivated QCD analysis of the behaviour
of the Adler $D$-function in the Euclidian region is described.
It is stressed that by taking account of $b$-quark
mass-dependent $\alpha_s^2$-effects one obtains better agreement
between theoretical predictions and experimentally motivated behaviour of
the $D$-function at large Euclidian momentum transfer.
A more detailed analysis of QCD predictions,
including information on quark and gluon condensates,
requires more precise data on
$e^+e^-\rightarrow{hadrons}$, particularly
in the energy regions $E<M_{J/\Psi}$
and $M_{J/\Psi}<E<3.6~{\rm GeV}$. Use of experimental
determination of the $D$-function to test the generalized
Crewther relation is outlined.
\end{abstract}
\maketitle

{\bf 1.~Introduction.}~
The process $e^+e^-$--annihilation into hadrons is one of
the most informative processes in elementary particle physics.
Over the past few decades special attention has been paid to
detailed theoretical and experimental study of its basic characteristics,
and in particular, of the ratio $R^{e^+e^-}(s)=\sigma_{tot}(e^+e^-\rightarrow
{hadrons})/\sigma(e^+e^-\rightarrow \mu^+\mu^-)$. As a result, important
information about the properties of hadrons and of their constituents
(quarks and gluons) was obtained. For example,
$e^+e^-$ collisions enabled
discovery in Novosibirsk of the first direct evidence for
the $\rho^0$-meson \cite{rho}. The $J/\Psi$ resonance \cite{J}
was simultaneously discovered by the $e^+e^-$-collider SPEAR at SLAC,
and at BNL as a result
of study of the process
$p+p\rightarrow e^+e^- + X$ \cite{Ting}. Rapid confirmation
came from a slight increase of the beam energy of
the ADONE $e^+e^-$ collider \cite{ADONE}.

The  observation
of $\Psi^{'}$ and $\Psi^{''}$ particles and their
interpretation  as $\overline{c}c$ bound states of quarks
(e.g. see the review \cite{ECHAYA}) has served as an
incitement
to
the development
of the theory of strong interactions and QCD in particular. It brought up the
question of the
possibility of  finding  the signal for the fifth quark in
$e^+e^-$-annihilation. Moreover, an
indirect determination of the mass of the sixth $t$-quark was made from
detailed fits of the experimental data of the high-energy
$e^+e^-$-colliders LEP and SLC  at the $Z^0$-pole  taking
into account the effects of its virtual
propagation (for a review see  \cite{virtop}).
Impressively, the extracted mass
of the $t$-quark turned out to be in
agreement with the its  direct ``measurement'', resulting
from subsequent discovery of the 6th quark at the Tevatron.

Among present theoretical studies of data from $e^+e^-$-colliders data
are attempts to find the
phenomenologically allowed window for the mass of the Standard Model Higgs
boson.
These studies are based on combined fits of data from the high-energy
colliders
LEP and SLC (for a review see \cite{Marciano})  taking into
account the effect of running of the inverse electromagnetic coupling constant
from its low-energy value $\alpha^{-1}(0)\approx 137.0356...$ to $M_Z$.
At this reference scale its value was determined  using a
compilation of the available $e^+e^-$-data \cite{EJ1}. The result of
Ref.\cite{EJ1}
$\alpha^{-1}(M_Z)=128.896(90)$  was recently updated
to give $\alpha^{-1}(M_Z)=128.913(35)$ \cite{EJ2} (for a discussion of
the fast developing   situation in this area see the review of
Ref.\cite{Jeg} and other related works on the subject \cite{refs}).
Further estimates of experimental and theoretical uncertainties
in this important quantity are on the agenda.
\begin{table*}[hbt]
\setlength{\tabcolsep}{0.5pc}
\newlength{\digitwidth} \settowidth{\digitwidth}{\rm 0}
\catcode`?=\active \def?{\kern\digitwidth}
\caption{Classification of constructed and planned $e^+e^-$-colliding
machines}
\label{tab:effluents}
\begin{tabular*}{\textwidth}{@{}l@{\extracolsep{\fill}}rrrrrr}
\hline
 Group &  Machine    &  Location  &  Start Data & Status in 1999    &
Beam Energy (b.e.)
Region \\
       &                &            &                 & June  &  $E_{b.e.}$ +
$E_{b.e.}$ [${\rm GeV}$] \\
\hline
      &   AdA  & Frascati &  1960  & -- &   \\
      &        & ORSAY  &  1961 & --   & 0.2 + 0.2 \\
      &  VEPP-2  & Novosibirsk     &  1966 & --  &0.2--0.55 + 0.2--0.55 \\
I   &   ACO & ORSAY & 1966  & --  & 0.2--0.55 + 0.2--0.55 \\
    &  ADONE &   Frascati & 1969 & -- & 0.7--1.55 + 0.7--1.55 \\
    &  VEPP-2M & Novosibirsk & 1974 & working  &0.2--0.67 + 0.2--0.67 \\
    &  DCI     &  ORSAY     & 1976 & --  &0.5--1.7 +0.5--1.77 \\
    &  $\phi$-factory &      &      &                   \\
    &  DA$\Phi$NE & Frascati & 1999 & launched   & 0.51+ 0.51 \\
\hline

    &   CEA & Cambridge (USA) & 1971 & -- &1.5--3.5 + 1.5--3.5 \\
    &   SPEAR & SLAC &  1972 & --  &1.2--4.2 + 1.2--4.2 \\
    &  DORIS & Hamburg & 1974 & -- &1--5.1 + 1--5.1 \\
II  &  VEPP-4 &  Novosibirsk  & 1979 & planned to restart & 1.5--5 + 1.5--5 \\
    &  CESR   & Cornell & 1979 & working  &3--8 + 3--8 \\
    &  BEPC  & Bejing & 1991 & working  &1--2.5 + 1--2.5\\
    &  $c$-$\tau$- factory & Bejing & $>$1999 & was planned & 1.5--3 + 1.5--3 \\
\hline
    &  B-factory  &        &          & asymmetric colliders &
\\
III    &  PEP-II  & SLAC &  1999 & launched  & 9 + 3.1 \\
        &  B-factory &   KEK &  1999 & launched  &    8 + 3.5 \\
\hline
    & PETRA &  Hamburg & 1978 & -- & 5--19 + 5--19 \\
IV  & PEP   &  Stanford & 1980 & -- &5--18 + 5--18 \\
    & TRISTAN & KEK &  1987 & --   &25--30 + 25--30 \\
\hline
    &  SLC   & SLAC & 1989 &  working &45--50 + 45--50 \\
V    & LEP-I  & CERN  &  1989 & -- & 45-47 + 45-47 \\
     & LEP-II & CERN  & 1995 & working & 65-100 + 65 --100 \\
\hline
    & TESLA  & DESY &   $>$ 2005 & linear colliders & 250-500 + 250-500\\
VI  & NLC &  SLAC &  $>$ 2005 & ( projects under  & 250-500 + 250-500\\
& JLC &  KEK  & $>$
 2005 & discussions) &  250-500 + 250--500\\
\hline
\end{tabular*}
\end{table*}

{\bf 2. Electron-positron annihilation: experiment    vs QCD
predictions.}~
From the perspective of the scanned energy regions, and the
physical problems under investigation,
all past, present and proposed $e^+e^-$-colliders may be divided
into six groups. We present this classification in Table 1,
updating the material
given in
Ref.\cite{GKL}. The new information was gained, in part,
from material in Ref.\cite{WT}.

These colliders provide complementary
information about the behaviour of the $e^+e^-\rightarrow{hadrons}$ total
cross-section,
at different energies, from machines of different luminosity.
Moreover,
some important physical problems under investigation necessitate
more precise experimental data, not only at high energies, but also
in low and
intermediate energy regions. Indeed, the study of the latter regions,
both by experimental and theoretical methods, may provide better
estimates of
\begin{equation}
\Delta\alpha_{had}(q^2)=-\frac{\alpha q^2}{3\pi} Re
\int_{4m_{\pi}^2}^{\infty}
ds\frac{R^{e^+e^-}(s)}{s(s-q^2-i\epsilon)}
\end{equation}
which is the hadronic vacuum-polarization contribution
to the value of the effective fine structure constant.

Another important characteristic of the process $e^+e^-$-hadrons
was introduced in Ref.\cite{Adler}. It concerns the Euclidian  
Adler $D$-function
\begin{equation}
D(Q^2)=Q^2\int_{4m_{\pi}^2}^{\infty}\frac{R^{e^+e^-}(s)}{(s+Q^2)^2}ds
\end{equation}
which can be related to Eq.(1) as follows:
\begin{equation}
D(-s)=\frac{3\pi}{\alpha}s\frac{d}{ds}\Delta\alpha_{had}(s) 
\end{equation}
The study of this quantity has a number of attractive features.
Its behaviour was first analyzed from an experimental point of view
some time ago \cite{Adler,RG}.
In the recent work of Ref.\cite{EJKV}
this problem was reconsidered using a
compilation of $e^+e^-$ experimental data
\cite{EJ1}\footnote{It would be interesting to include in this
and related analyses the new measurements of
$R^{e^+e^-}(s)$ at $\sqrt{s}$ values of 2.6, 3.2, 3.4, 3.55, 4.6 and 5 ${\rm GeV}$,
from BEPC \cite{new}.}.
The results are depicted in Fig.1 and Fig.2 where the shaded
areas represent the $\pm1\sigma$ band obtained from the data.
\\[-2.8cm]
\begin{figure}[htb]
\medskip
\parbox[b]{7cm}{\psfig{width=6cm,angle=90,file=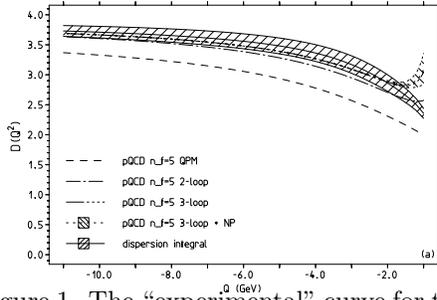}}
\vspace{-1.8cm}
\caption{The ``experimental'' curve for the Adler function
together with QCD  predictions}
\end{figure}
\\[-2.8cm]
\begin{figure}[htb]
\vspace{-25pt}
\medskip
\parbox[b]{7cm}{\psfig{width=6cm,angle=90,file=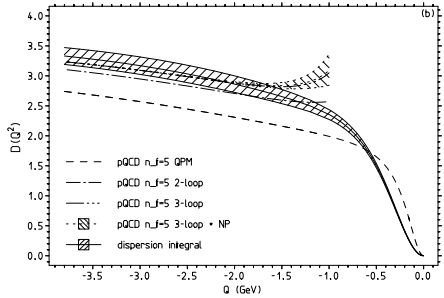}}
\vspace{-1.5cm}
\caption{The low energy  ``experimental'' and QCD
behaviour of the Adler function}
\end{figure}
\\[-1.0cm]
It is interesting to compare the  ``experimental'' behaviour for the
$D$-function with its QCD theoretical expression.
One can express it in the following form
\begin{equation}
D^{QCD}(Q^2)=D^{PT}(Q^2)+D^{NP}(Q^2)
\end{equation}
where the first part describes  the perturbative QCD contributions,
while the second term takes account of higher-twist
effects, which are related  to the vacuum condensates of quark and
gluon fields \cite{ShVZ}. In the work of Ref.\cite{EJKV} we considered
an $\alpha_s^2$ mass-dependent expression for $D^{PT}(Q^2)$, namely
\begin{eqnarray}
D^{PT}(Q^2)&=&D^{(0)}(Q^2)+D^{(1)}(Q^2)\frac{\alpha_s(Q^2)}{\pi} \\ \nonumber
&+&
D^{(2)}(Q^2)(\frac{\alpha_s(Q^2)}{\pi})^2
\end{eqnarray}
The Born term expression is well known and has
the following form
\begin{eqnarray}
D^{(0)}(Q^2)&=& 3\sum_f Q_f^2\bigg(1-6x \\ \nonumber
&+&\frac{12x^2}{\sqrt{1+4x^2}}\ln
\bigg[\frac{\sqrt{1+4x}+1}{\sqrt{1+4x}-1}\bigg]\bigg)
\end{eqnarray}
where $x=m_f^2/Q^2$, while $m_f$ is the pole quark mass.
The analytic mass-dependent expression for $D^{(1)}$
can be obtained  from the calculations of Ref.\cite{FJ},
which are in agreement
with the results of
Ref.\cite{oth}.
Since the mass dependence of the 3-loop term
$D^{(2)}$-term in Eq.(5)   is not yet known
analytically,
detailed information about its   small-mass
\cite{GKL2,Chetetal} and  heavy-mass \cite{ChKS} expansions
was used in
Ref.\cite{EJKV} to approximate the behaviour of $D^{(2)}$ in the
Euclidian region,
with high  precision, using conformal mapping
and Pad\'e improvement \cite{T}. However, the theoretical expression
for mass dependence of the double-bubble three-loop vacuum polarization
diagram was not included in the theoretical expression for $D^{(2)}$-term
in Ref.\cite{EJKV}.
There are  arguments, based on the calculations of
$(m^2/Q^2)\alpha_s^2$-corrections \cite{GKL2}, that this contribution is
small.
\begin{table*}[hbt]
\setlength{\tabcolsep}{0.5pc}
\caption{ The origin of uncertainties for the ``experimental'' Adler
function at $Q=2.5~{\rm GeV}$ and $M_Z$
}
\label{tab:effluents}
\begin{tabular*}{\textwidth}{@{}l@{\extracolsep{\fill}}rrrrrr}
\hline
    &  $D(2.5~{\rm GeV}$) & rel. err. & abs. err.     &  $D(M_Z)$  &  rel. err.  &
abs. err  \\
\hline
Resonances      &  .688 (.025)  & 3.6 $\%$ & 0.8 $\%$  & .004 (.000) &
5.2 $\%$ & 0.0 $\%$ \\
 $E<M_{J/\Psi}$ & 1.068 (.127) & 11.9 $\%$ & 4.2 $\%$  & .002 (.000) &
14.9  $\%$ & 0.0 $\%$   \\
 $M_{J/\Psi}<E<3.6~{\rm GeV}$    &  .178 (.035)  & 19.9 $\%$ &  1.2 $\%$ &
 .001 (.000) & 19.8 $\%$ & 0.0 $\%$  \\
$3.6~{\rm GeV}<E<M_{\Upsilon}$    & .850 (.055) & 6.4 $\%$  & 1.8 $\%$ & .032 (.002) &
7.0 $\%$ & 0.1 $\%$  \\
$M_{\Upsilon}<E<12~{\rm GeV}$     &  .088 (.008) &  8.7 $\%$  & 0.3 $\%$ & .024 (.002) &
9.0 $\%$ & 0.1 $\%$ \\
$E<12~{\rm GeV}$ data    &  2.871 (.146)  & 5.1 $\%$  & 4.8 $\%$  & .063 (.003) &
4.9 $\%$ & 0.1 $\%$ \\
$12~{\rm GeV} < E$ QCD & .162 (.001) & 0.3 $\%$ & 0.0 $\%$ & 3.755 (.006) & 0.2 $\%$ &
0.2 $\%$ \\
total & 3.033 (.146) & 4.8 $\%$ & 4.8 $\%$ & 3.818 (.007) & 0.2 $\%$ & 0.2 $\%$ \\
\hline
\end{tabular*}
\end{table*}
Let us consider the non-perturbative contributions
to $D^{NP}(Q^2)$. In Ref.\cite{EJKV} the following expression was used
\begin{eqnarray}
D^{NP}(Q^2)&=& 3\sum_f Q_f^2 (8\pi^2) \\ \nonumber
&\times&\bigg[\frac{1}{12}\bigg(1-\frac{11}{18}
\frac{\alpha_s(\mu^2)}{\pi}\bigg)\frac{<\frac{\alpha_s}{\pi}G^2>}{Q^2}
\\ \nonumber
&+& 2 \bigg(1+\frac{\alpha_s(\mu^2)}{3\pi}
+ O(\alpha_s^2)\bigg)\frac{<m_q\overline{q}q>}{Q^4} \\ \nonumber
&+&\bigg(\frac{4}{27}\frac{\alpha_s(\mu^2)}{\pi}+O(\alpha_s^2)\bigg)
\sum_{q^{'}}
\frac{<m_{q^{'}}\overline{q^{'}}q^{'}>}{Q^4}
\end{eqnarray}
The terms beyond leading order in $\alpha_s$, calculated in the works of
Ref.\cite{power}, turned out to be not so important at the moment, because
the current experimental data for the $D$-function gives only rough
constraints on
the contributions of dimension-four condensates, which were varied in
Ref.\cite{EJKV} within the
following intervals $<\frac{\alpha_s}{\pi}G^2>\approx (0.336-0.442~{\rm GeV})^4$,
$<m_u\overline{u}u>=<m_d\overline{d}d>=-(0.086-0.111~{\rm GeV})^4$,
$<m_s\overline{s}s>=-(0.192-0.245~{\rm GeV})^4$.
Nevertheless, in Fig.1 and Fig.2 one can see the region, where the
addition of the nonperturbative QCD corrections to   the three-loop
perturbative QCD expression leads to deviation from
the experimentally
allowed region
for the $D$-function at
low Euclidian momentum transfer, $Q^2$.
It is  possible, that
taking into account  higher-order power corrections \cite{highp} might improve
the agreement with ``experimental data''  at low energies, shown in Fig.2.
Moreover, it is of real interest to update the previous
analysis, in Refs.\cite{EKV,GP},
of the low-energy $e^+e^-$ experimental data ( see
Refs.\cite{EKV,GP}) with the help of the
QCD Borel sum rule method \cite{ShVZ}
\begin{equation}
\int_0^{\infty} R^{th}(s)e^{-s/M^2}ds =
\int_0^{\infty} R^{e^+e^-}(s) e^{-s/M^2}ds~~~~~~~.
\end{equation}
In the process of such an
analysis,  new low-energy experimental data,
obtained in part in Novosibirsk, can be used. In the theoretical part
of sum rule of Eq.(8)
one should also include
perturbative contributions to the
coefficients function of quark and gluon condensates  of dimension four
\cite{power}, available from the results
of Ref.\cite{highp}, higher-dimension condensates,
and information about massless
$\alpha_s^3$
contributions to $R^{th}(s)$ \cite{GKLR}. As to the sum-rule analysis
of the high-energy $e^+e^-$ experimental data, one can
try to update the studies of Ref.\cite{FESR}, performed
with the help of the finite energy
sum rules
approach
\begin{equation}
\int_0^{s_0}R^{th}(s)ds = \int_0^{s_0} R^{e^+e^-}(s) ds~~~~~~~~.
\end{equation}

Concerning the
curves for
the $D$-function
extracted from experimental data
(see Figs. 1,2 of Ref.\cite{EJKV}), it is worth noting
that the theoretical analyses of the recent works \cite{ShS,IG} result
in a description of the
low-energy  tail
of Fig.2
by completely
different ideas, related to the concept  of ``freezing'' of the QCD
coupling constant at small energies \cite{ShS1,DMW}.
These ideas have realizations,
distinct  from conclusions in Ref.\cite{MS},
based on application of the PMS approach \cite{PMS} to the
four-loop massless theoretical expression for $R^{e^+e^-}(s)$.
Indeed, it was  argued in Ref.\cite{ChKL}, that the observed ``perturbative''
freezing may be spurious, indicating breakdown
of a next-to-next-to-leading PMS expression, around
the  scale
of $\rho$-meson mass. Note, however, that the low-energy results
for the  fit of
$e^+e^-$ data on $R^{e^+e^-}(s)$, performed in Ref.\cite{MS} with the
help of the following  sum rule \cite{PQW}
\begin{equation}
\overline{R}(s,\Delta) =
\frac{\Delta}{\pi}\int_0^{\infty}\frac{R^{e^+e^-}(s^{'})}{(s^{'}-s)^2+
\Delta^2}ds^{'}~~~~~,
\end{equation}
merited serious attention. It should be stressed that the low-energy
Novosibirsk data of Ref.\cite{Barkov} turned out to be essential in
this analysis. For example, they served as an  ingredient in the
phenomenological part of the considerations of Ref.\cite{ShS2}, devoted
to the analysis of ``analytical'' freezing of the QCD coupling constant
$\alpha_s$ in the Minkowskian region, and also in Ref.\cite{BGKL}, devoted
to consideration of the Crewther relation
\cite{Crewther} and its $\overline{MS}$-scheme QCD generalization \cite{BK},
using commensurate scale relations \cite{BL} (for
reviews see \cite{Kataev,Brodsky}). It should be recalled that  the
$\alpha_s^3$-generalization of the  Crewther relation connects
a massless
$\alpha_s^3$
theoretical
expression for the $e^+e^-$-annihilation $D$-function \cite{GKLR} with
the massless theoretical expressions for the
Gross-Llewellyn Smith sum rule of the $\nu N$ deep-inelastic scattering
and the Bjorken sum rule of  polarized deep-inelastic scattering,
which were
calculated at the $\alpha_s^3$-level in Ref.\cite{LV}. This connection
involves the first and second terms
of the two-loop approximation to the QCD $\beta$-function. We
will
consider  this  problem in more detail in the next Section.

We now return to the results of Ref.\cite{EJKV}.
Before discussing   the comparison of the three-loop
massive theoretical QCD predictions for the $D$-function with
the ``experimental'' behaviour of the $D$-function at higher
momentum transfer,
it is worth mentioning that in the process of extracting
the ``experimental'' behaviour of the $D$-function not all $e^+e^-$
data up to
$40~{\rm GeV}$
were applied. Indeed,  in Ref.\cite{EJKV} a conservative attitude was
adopted:
the real data were replaced by perturbative QCD results in certain regions,
but only where it was obviously
safe to do so. Thus,  in the regions from $4.5~{\rm GeV}$ to $M_{\Upsilon}$ and above
$12~{\rm GeV}$ perturbative QCD results were used,
including massive three-loop \cite{Chetetal} and a massless
four-loop QCD contribution \cite{GKLR}.

The origins of uncertainty in the
``experimental''
curves for the $D$-function were analyzed in Ref.\cite{EJKV}
(see  Table 2).  The main sources
come from the region $E<M_{J/\Psi}$, {\bf accessible for more detailed
experimental
inspection at VEPP-2M and  BEPC}, and the region $M_{J/\Psi}<E<3.6~{\rm GeV}$,
{\bf which is the privilege of BEPC, VEPP-4}  and a possible future $c-\tau$
factory.
However, as was shown in Ref.\cite{EJKV}, even at the current level of
experimental precision one can already obtain new  information, namely
a demonstration of the importance of
two-loop heavy-quark mass-dependent
effects for the
``experimental'' behaviour of the $D$-function at
moderate and high Euclidian momentum transfers.
Indeed,  after including mass effects, both in the
three-loop perturbative
part of the $D$-function,
and also in the two-loop  running of the QCD coupling
constant
$\alpha_s$ from the value of $\alpha_{s,\overline{MS}}(M_Z)=0.120 \pm 0.003$
to lower energy scales, via a
variant in Ref.\cite{ET} of the
momentum~\footnote{For earlier discussion
of the advantages of MOM-type schemes, taking
account of threshold effects, see
Ref.\cite{Shirkov}}
subtraction scheme,
one can observe the appearance of real agreement
of the three-loop  massive theoretical expression
for the $D$-function with the  experimentally-motivated
Euclidian curves of Fig.1
and Fig.2. It should be stressed, that if we had not included the three-loop
massive term, a discrepancy with the
``experimental'' behaviour of the $D$-function
might have been interpreted as
requiring non-perturbative power corrections
from Eq.(7). While the addition of the twist-4 power
corrections can be of real importance in the region of small
enough $Q^2$, deviation of the two-loop curves
from the ``experimental'' corridor for the $D$-function at high
momentum transfer (see Fig.1) can be associated with omission of
perturbative QCD contributions.
Another interesting observation is that
the curves of Fig.1 and Fig.2
turned out to be rather smooth \cite{EJKV}, lacking the resonance
enhancements and ``threshold steps'',
typical of the Minkowskian region. In view of this we think that
possible future applications and improvements
of the
results obtained in
Ref.\cite{EJKV}
can be useful for more detailed tests of perturbative and
non-perturbative theoretical QCD predictions in the Euclidian region.
One such application is presented in the next Section.

{\bf 3. New tests of the generalized Crewther relation.}~
The Crewther relation \cite{Crewther} connects
the amplitude
of $\pi^0\rightarrow\gamma\gamma$ decay with
the product of the $e^+e^-$ annihilation $D$-function and
deep-inelastic scattering
sum rules, namely with
Bjorken sum rule of polarized lepton-nucleon scattering
\begin{eqnarray}
Bjp(Q^2)&=&
\frac{1}{6}|\frac{g_A}{g_V}|C_{Bjp}(Q^2) \\ \nonumber
&=&
\int_0^1[g_1^{ep}(x,Q^2)-g_1^{en}(x,Q^2)]dx
\end{eqnarray}
or with the Gross-Llewellyn Smith sum rule of $\nu N$
deep-inelastic scattering
\begin{equation}
GLS(Q^2)=3C_{GLS}(Q^2)=\int_0^1F_3(x,Q^2)dx
\end{equation}
where the analytical massless perturbative theory expression for the
$D(Q^2)$-function is known at the $\alpha_s^3$-level \cite{GKLR},
while the massless
analytical perturbative theory expressions for
$C_{Bjp}(Q^2)$ and $C_{GLS}(Q^2)$ up to $\alpha_s^3$-corrections are known
from the calculations of Ref.\cite{LV}. It should be stressed that all the
quantities
we will be interested in are defined in the Euclidian region and
all the results
were obtained in the $\overline{MS}$-scheme. In this scheme the
$\alpha_s^3$ generalization of the Crewther relation,
discovered in Ref.\cite{BK}, has the
following form:
\begin{equation}
C_{Bjp}(\alpha_s(Q^2))C_{D}(\alpha_s(Q^2))=
1+\frac{\beta^{(2)}(\alpha_s)}{\alpha_s}P(\alpha_s)
\end{equation}
where $C_{D}(\alpha_s(Q^2))$ is
coefficient function
for the Adler $D$-function,
normalized to unity at lowest order,
$\beta^{(2)}(\alpha_s)$ is the two-loop
approximation
of the QCD $\beta$-function and $P(\alpha_s)$ is a polynomial
starts
from $\alpha_s$ and  contains two terms. It was suggested
in Refs.\cite{BK,GK} that
the factorization of the $\beta$-function will persist generally,
to all perturbative orders, and can be related to the effects of
violation of
conformal symmetry by the renormalization of massless QCD.
This was later proved in
Ref.\cite{Crewther2}. The theoretical properties of the
generalized
relation, written down in the $\overline{MS}$-scheme,  were discussed in
detail in Refs.\cite{BK,Kataev,Crewther2} and we will avoid their
description in this work. However, we will concentrate
on some phenomenological applications.  It is rather useful
to use for this purpose commensurate scale relations
\cite{BL}, which combine the concept of
effective charges  \cite{Grunberg} (or scheme-invariant perturbation
theory \cite{DG}) with variants of the BLM approach \cite{BLM,GrK},
allowing one
to write the following generalization of the Crewther
relation \cite{BGKL} in the region where the heavy-quark masses can
be neglected:
\begin{eqnarray}
\frac{1}{3\sum_f Q_f^2}D(Q^2)C_{GLS}(Q_*^2)\approx 1 \\ \nonumber
\frac{1}{3\sum_f Q_f^2}D(Q^2)C_{Bjp}(Q_*^2) \approx 1
\end{eqnarray}
where
\begin{eqnarray}
\ln \frac{Q_*^2}{Q^2}&=&-\frac{7}{2}+4\zeta_3 \\
\nonumber
&&
+\bigg(\frac{\alpha_{GLS}(Q_*^2)}{4\pi}\bigg)\bigg[\big(\frac{11}{12}+
\frac{56}{3}\zeta_3-16\zeta^2_3\big)\beta_0 \\ \nonumber
&&-\frac{56}{27}-\frac{808}{9}\zeta_3
+\frac{320}{3}\zeta_5\bigg]
\end{eqnarray}

As specimen physical input we will first use the recently obtained
value of the Gross-Llewellyn Smith sum rule \cite{GLS}
\begin{equation}
GLS(12.59~{\rm GeV}^2)=2.80 \pm 0.13 \pm 0.17~~~~~
\end{equation}
which does not contradict  the less precise similar
results, obtained using the extrapolation of the
data \cite{KatSid}.
Using the results  of Ref.\cite{CK} we conclude that, in this  energy
region, $f=4$ numbers of flavours are manifesting themselves. Moreover,
at momentum transfer $Q_*^2=12.59~{\rm GeV}^2$ we will neglect
$c$-quark mass
effects, which are
suppressed by a factor $m_c^2/Q_*^2<0.19$. Thus
we conclude
that the value of the corresponding effective charge is
\begin{eqnarray}
\frac{\alpha_{GLS}(Q_*^2)}{\pi}&=&1- \frac{GLS(Q_*^2)}{3}  \\
\nonumber
&=& 0.067 \pm 0.043 \pm 0.06
\end{eqnarray}
where the first (second) error is related to the statistical
(systematical) uncertainty of the experimental result
of Eq.(16).
Using Eqs.(14)-(17) one can get the following estimate
\begin{equation}
D(Q\approx 1.8~{\rm GeV}) \approx 3.57 \pm 0.3
\end{equation}
which crosses the upper part of
``experimentally'' motivated curve of
Fig.2 at this
reference scale. A similar conclusion emerges
in the case of analogous
treatment of the experimental result of the SMC collaboration for the
polarized Bjorken sum rule \cite{SMC}, which is
\begin{eqnarray}
Bjp(Q^2=10~{\rm GeV}^2)&=&0.195 \pm 0.029 \\ \nonumber
&=&\frac{1}{6}|\frac{g_A}{g_V}|
\bigg(1-\frac{\alpha_{Bjp}(Q_*^2)}{\pi}\bigg)
\end{eqnarray}
As a result of application of discussions in
Ref.\cite{MSS} and considerations presented above,
we obtain
the following estimate for the $D$-function
\begin{equation}
D(Q\approx 1.6~{\rm GeV})\approx 3.59 ^{+0.35}_{-0.45}
\end{equation}
which is  in rough agreement with the ``experimentally'' motivated
Euclidian
results of Ref.\cite{EJKV}, though with larger errors.

Of course, our considerations are not so detailed as the ones
given in Ref.\cite{ABFR}, where an expression for the value of the
coupling constant $\alpha_s$ was extracted from the Bjorken sum rule data
(for earlier discussions of this problem see Ref.\cite{Ellis}).
However, we believe that even approximate tests of the generalized
Crewther relation, written down
in the form of commensurate scale
relations of Ref.\cite{BGKL} {\bf directly in the Euclidian region},
give one the feeling that
estimates  obtained from  deep-inelastic sum rules are less
precise
than direct extraction of the behaviour of the $D$-function in the
Euclidian region. This might suggest a need for adding
higher-twist contributions to
the perturbative generalization of the Crewther relation.

{\bf Acknowledgements.}
We are grateful to the organizers of the International Workshop
``$E^+E^-$-Collisions from $\phi$ to $J/\Psi$'' (Novosibirsk, March 1-5,
1999)
for the invitation, hospitality and financial support.
It is a pleasure to thank S.I. Eidelman, F. Jegerlehner and O.L. Veretin
for the fruitful and pleasant collaboration, which resulted
in studies  of the experimentally motivated  and
theoretical behaviour of the Adler $D$-function and S.J. Brodsky,
G.T. Gabadadze, H.J. Lu and especially D.J. Broadhurst for sharing
their insight during the common derivations of the various forms of
the generalized Crewther
relation. I am also grateful to D.J. Broadhurst for making 
some style corrections in the previous draft of this work.
This work was written during a stay at the Abdus Salam International
Centre for Theoretical Physics, Trieste, Italy.
I would like to thank my colleagues from ICTP for hospitality.
I also wish to thank I.A. Savin and D.V. Peshekhonov for discussions
on the possibility of different applications of the polarized Bjorken sum
rules data of the SMC Collaboration.

\end{document}